\documentclass{PoS}

\title{Study of finite temperature QCD with 2+1 flavors via Taylor expansion and imaginary chemical potential}

\ShortTitle{QCD with 2+1 flavors via Taylor expansion
and imaginary chemical potential}

\author{\speaker{Rossella Falcone}\\
        Fakult\"{a}t f\"{u}r Physik, Universit\"{a}t Bielefeld, D-33615 Bielefeld, Germany\\
        E-mail: \email{rfalcone@physik.uni-bielefeld.de}}

\author{Edwin Laermann\\
        Fakult\"{a}t f\"{u}r Physik, Universit\"{a}t Bielefeld, D-33615 Bielefeld, Germany\\
        E-mail: \email{edwin@physik.uni-bielefeld.de}}

\author{Maria Paola Lombardo\\
        INFN - Laboratori Nazionali di Frascati, I-00044 Frascati (RM), Italy\\
        E-mail: \email{Mariapaola.Lombardo@lnf.infn.it}}

\abstract{We study QCD with 2+1 flavors at nonzero temperature and nonzero 
chemical potential. We present preliminary results obtained from lattice 
calculations performed with an improved staggered fermions action (p4-action) 
on lattice with temporal extent $N_t = 4$ on a line of constant physics 
with the strange quark mass adjusted to its physical value and a pion 
mass of about 220 MeV. We compute at imaginary chemical potential and 
compare with Taylor expansion results. We focus our study on a range of temperatures 
$0.94 < T/T_c < 1.08$.}

\FullConference{The XXVIII International Symposium on Lattice Field Theory, Lattice2010\\
		June 14-19, 2010\\
		Villasimius, Italy}

\begin{document}

\section{Introduction}
Understanding the phase diagram of QCD in the temperature - chemical 
potential (T,$\mu$) plane is crucial for many implications in astrophysics,
cosmology and in the phenomenology of heavy-ion collisions.
While the lattice formulation of QCD provided fruitful information at finite 
temperature, at non-vanishing chemical potential lattice simulations suffer 
from the sign problem. Only in recent years 
several different methods 
\cite{Barbour,Fodor,deForcrand,DElia,Gavai,Allton,deForcrand2}
have been devised to at least partly overcome this obstacle.

In this work we present results from lattice calculations
in QCD with dynamical light and strange quark degrees 
of freedom at non zero temperature and imaginary chemical potential. 
Our calculations are performed with a tree level Symanzik-improved 
gauge action and an improved staggered fermion action, the p4-action with 
3-link smearing (p4fat3)~\cite{Karsch_p4fat3}.
We focus on a range of temperatures in the vicinity of the pseudocritical
temperature $T_c$ at vanishing chemical potential,
$0.94 < T/T_c < 1.08$. At each temperature
we carried out simulations on lattices with temporal extent $N_t = 4$.
Following~\cite{Cheng}, at each temperature
the strange quark mass was adjusted to its physical value and 
the light up and down quark masses were taken to be degenerate and equal
to $m_s/10$, which corresponds to a constant Goldstone pion mass of
about 220 MeV.
This allows us to utilize the zero temperature results and interpolations
of~\cite{Cheng} for the conversion of lattice parameters into physical
units.
For definiteness, we apply a value of $T_c = 202$~MeV for the pseudocritical
temperature at zero density.
To insure small finite volume effects
the spatial volume has been chosen to be $V^{1/3}T=4$. 
The number of gauge field configurations analyzed varies from 1000
to 2000, the configurations are separated by 10 trajectories.
All numerical simulations have been performed using the Rational Hybrid
Monte Carlo (RHMC) algorithm~\cite{rhmc,p4_Tc_nf3}.

At each temperature we performed calculations at several values of the
imaginary quark chemical potential, $\mu_q = i \mu_I, q= u,d,s$ which
was taken to be degenerate for all three flavors. In the following, 
we first present results relevant for an estimate of the pseudocritical
line in the $(T,\mu_I)$ plane and then compare our findings at imaginary
chemical potential with results obtained within the Taylor expansion
approach.


\section{The pseudocritical line at imaginary chemical-potential}

At imaginary chemical potential and high temperature,
the phase diagram of QCD 
features  
the Roberge Weiss transition at $\mu^c_I=\pi T/3$, associated with the phase
of the Polyakov loop. 
The line of Roberge Weiss transitions ends at $T = T_{RW}$. 
At lower temperature the `quark-gluon plasma' region is limited by a 
chiral transition~\cite{deForcrand,DElia} which continues to
real values of $\mu$.
In this work we focus our study on the range of temperatures 
$0.94 < T/T_c < 1.08$, i.e. to the vicinity of the pseudocritical
temperature $T_c$ at zero density.

\begin{figure}[t]
  \centering
  \bigskip
  \includegraphics[width=.55\textwidth]{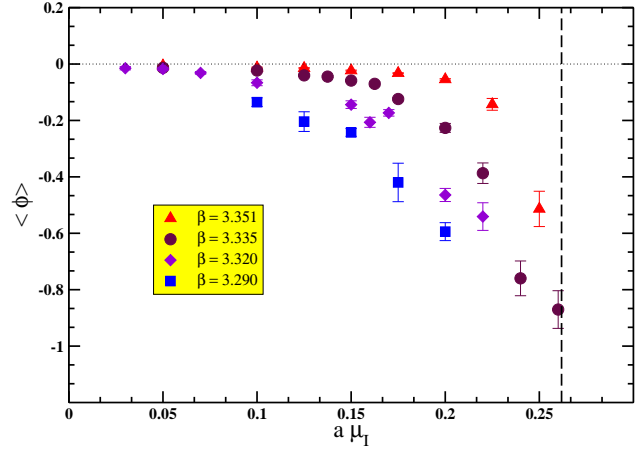}
  \caption{The Polyakov loop phase $\langle\phi\rangle$ as a function
    of the imaginary chemical potential for 4 different beta values. 
    The vertical dashed line corresponds to the RW transition
    at $a \mu_I^c = \pi/3 N_t$.}
  \label{fig:phase}
\end{figure}

In order to study the phase structure of QCD in the $(T,\mu_I)$ plane,
it is very useful~\cite{DElia} to consider the phase of the 
Polyakov 
loop, $L(x)$, that we can parameterize as $L(x)=|L(x)|e^{i\phi}$.
In the presence of dynamical fermions and with imaginary chemical potential
we expect $\langle\phi\rangle = - \theta$ at low temperatures 
($\theta\equiv \mu_I/T$),
and
$\langle\phi\rangle \sim 2k\pi/3$ for 
$(k - 1/2) < \frac{3}{2\pi}\theta < (k + 1/2)$ 
at high temperatures; the values $\theta=2(k + 1/2)\pi/3$ correspond
to the Roberge Weiss transitions from one $Z_3$ sector to the other.
In Fig.~\ref{fig:phase} we show our results for $\langle\phi\rangle$
versus the imaginary chemical potential for different values of temperatures.
For $\beta=3.290$ which is below the critical value at $\mu_I=0$ 
we found that 
$\langle\phi\rangle$ starts to deviate from zero right away at
non vanishing potential. 
For the $\beta$ values above the critical one we observe
that $\langle\phi\rangle$ stays close to 0 until a certain
temperature dependent value of $a\mu_I$
is reached where the Polyakov loop phase begins to decrease with
increasing $\mu_I$. 
We will take that value as an estimate for the location
of the pseudocritical chiral line~\cite{DElia}.
A jump of the phase is not observed at any temperature
which indicates that all our $T$ values are below $T_{RW}$

\begin{figure}[htb]
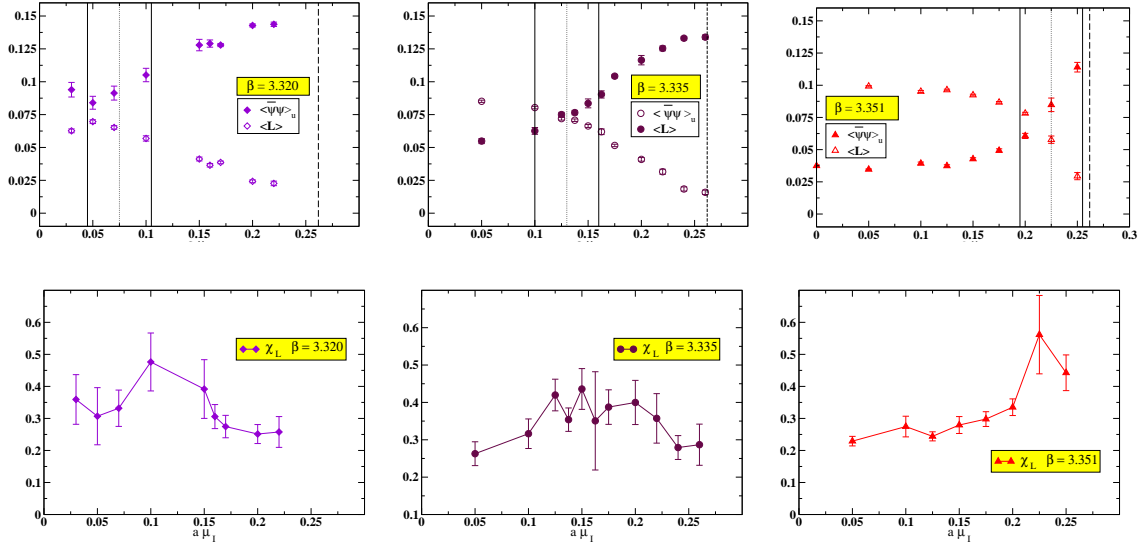

  \centering
  \bigskip
  \includegraphics[width=.31\textwidth]{chir_b3_320.eps}$\quad$
  \includegraphics[width=.31\textwidth]{chir_b3_335.eps}$\quad$
  \includegraphics[width=.31\textwidth]{chir_b3_351.eps}\\
  \bigskip
  \includegraphics[width=.3\textwidth]{pol_susc_b3.320.eps}$\quad$
  \includegraphics[width=.3\textwidth]{pol_susc_b3.335.eps}$\quad$
  \includegraphics[width=.3\textwidth]{pol_susc_b3.351.eps}
  \caption{Chiral condensate and Polyakov loop 
as functions of the imaginary chemical potential (upper row) for beta values
above the critical value. The vertical dashed line corresponds to the
RW transition, the vertical continuous ones to the estimation of the 
critical $\mu_I$ from the Polyakov loop phase; 
(lower row) Polyakov loop susceptibility as function
of $\mu_I$ for the same beta values. }
  \label{fig:pol&chir_susc}
\end{figure}

In Fig.~\ref{fig:pol&chir_susc} we display our results for the modulus
of the Polyakov loop
and the light quark chiral condensate as functions of the imaginary chemical
potential for $\beta=3.320,3.335,3.351$ which correspond 
to temperatures of 204.8 MeV, 209.6 MeV and 218.2 MeV
respectively \cite{Cheng}. The chiral condensate 
is defined as
\begin{equation}
  \langle \hat{\psi}\psi \rangle_{q} \equiv \frac{1}{4}\frac{1}{N_\sigma^3 N_t}\langle Tr M^{-1}({m}_q)\rangle,\quad q=u,d,s 
\end{equation}
where $N_\sigma^3$ is the spatial volume and $M$ the fermionic matrix.
In the same figure we also display the behavior of the Polyakov loop 
susceptibility
\begin{equation}
  \chi_L = N_\sigma^3\left(\langle L^2 \rangle - \langle L \rangle ^2\right)
\end{equation}
as a function of $\mu_I$, for the same beta values.
In the upper row of Fig.~\ref{fig:pol&chir_susc} we indicate
by vertical lines the estimates for the critical $\mu_I$ values and their
errors as obtained from the phase of the Polyakov loop.
It is seen, most clearly at the largest temperature investigated, that
in the region where the Polyakov phase starts to deviate from 0 also
the chiral condensate exhibits a rise while the modulus
of the Polyakov loop decreases.


\begin{figure}[htb]
  \centering
  \bigskip
  \includegraphics[width=5.5cm,height=4.5cm]{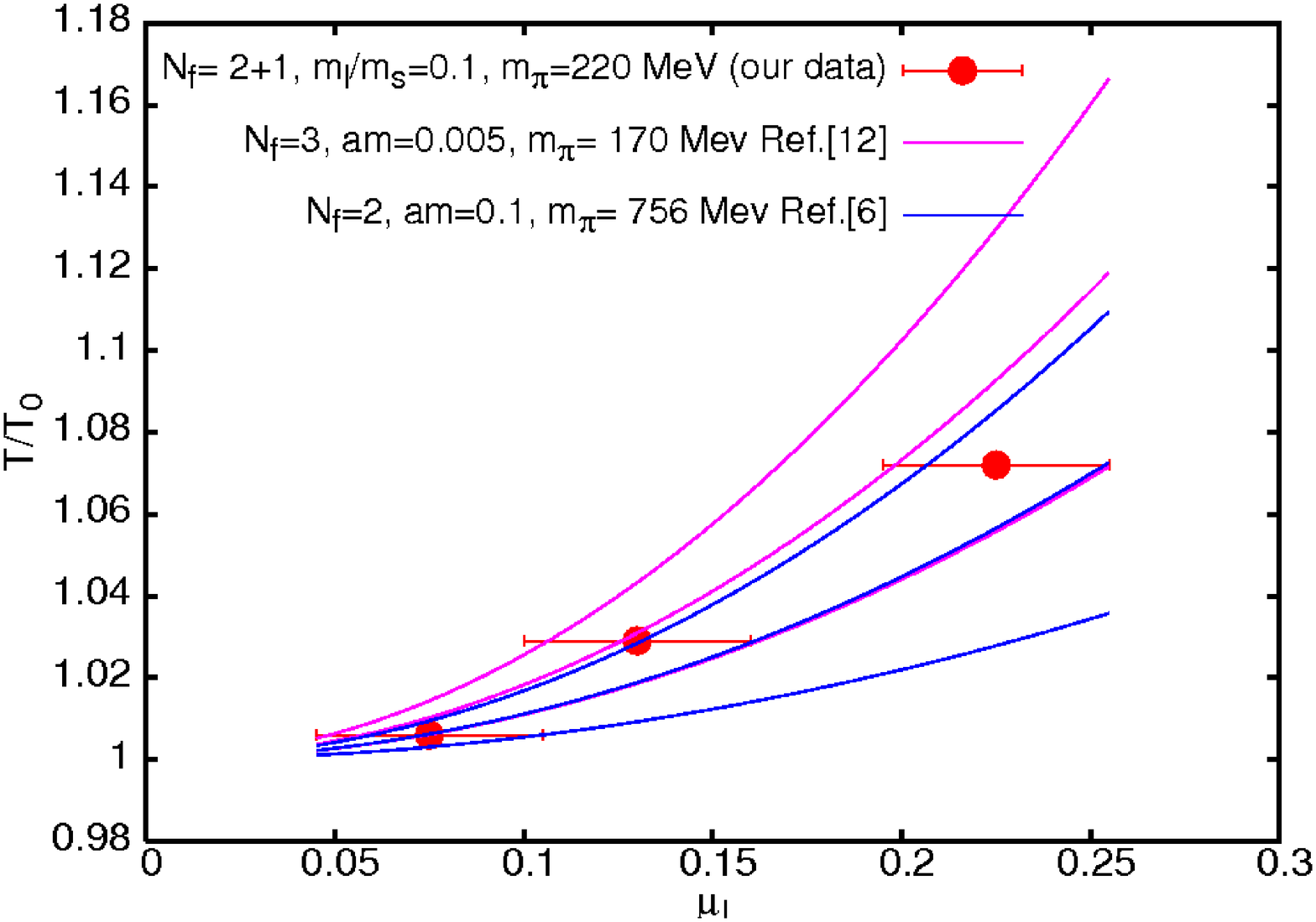}
  \includegraphics[width=5.5cm,height=4.5cm]{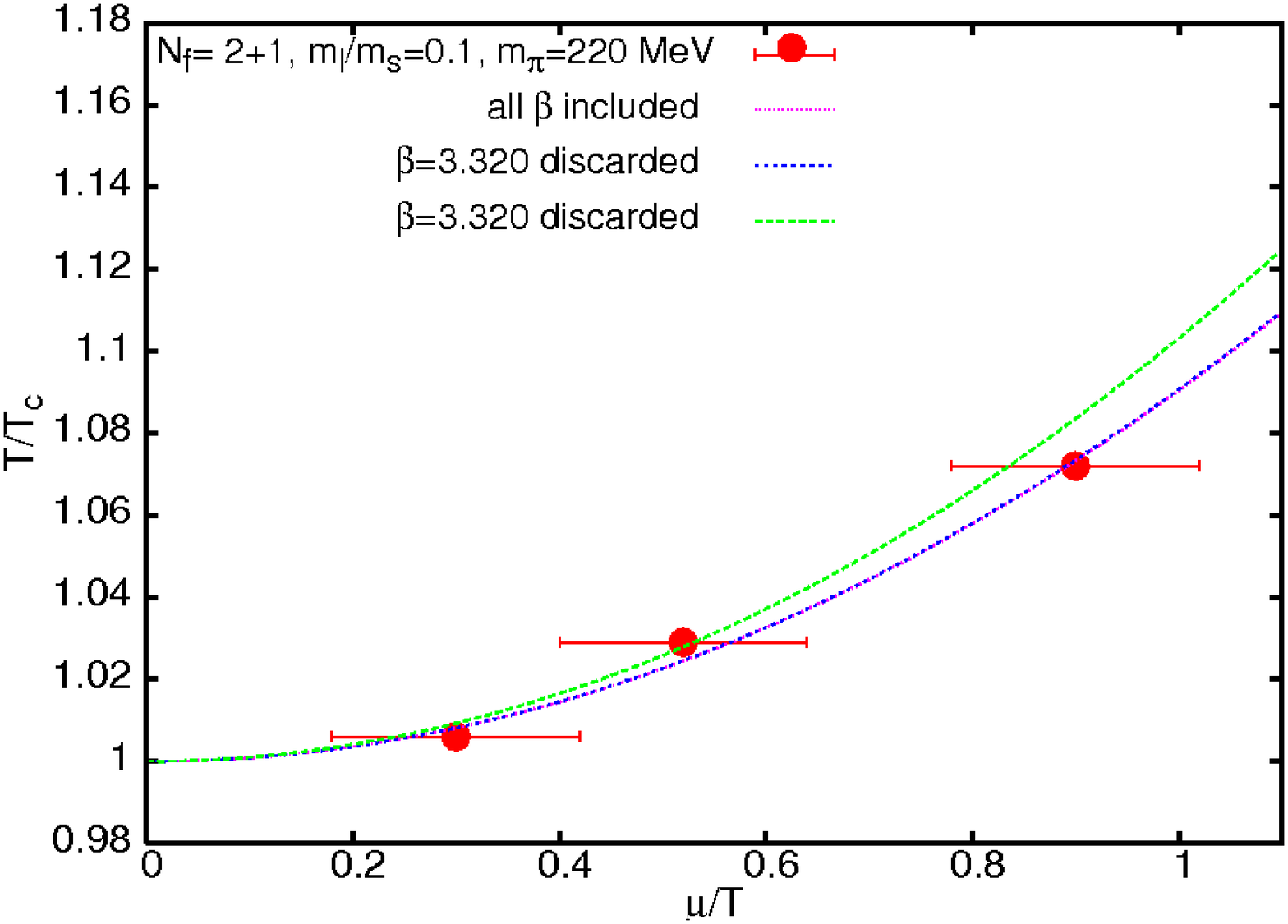}
  \caption{The pseudocritical line at imaginary $\mu$: comparison between our data points 
and results from the literature~\cite{Allton,Allton3} (left); our fit results (right).}
  \label{fig:critical_line}
\end{figure}

Based on the estimates for the pseudocritical $\mu_I$ values, 
to leading order in $\mu/T$
the critical line can be parameterized as
\begin{equation}
  \frac{T_c(\mu)}{T_c}=1-t_2(N_f,m_f)\left(\frac{\mu}{\pi T}\right)^2
  +{\mathcal O}\left( \left( \frac{\mu}{\pi T}\right)^4 \right)
\end{equation}
where $T_c$ is the critical temperature at zero chemical potential.
The coefficient of the leading term, $t_2$, has been calculated
for various cases, see the collection in~\cite{Schmidt2}. In 
Fig.~\ref{fig:critical_line} (left) we compare our data points on
$T_c(\mu)/T_c$ with results which have been obtained from
reweighting simulations at $\mu=0$ within the same discretization
scheme as ours. In~\cite{Allton} two rather heavy quark
flavors were taken into account whereas in~\cite{Allton3}
the number of flavors was 3, with similar masses as in our case.
Although the error bars are large the comparison suggests that the
curvature $t_2$ grows with $N_f$ which
 is consistent with a behavior $\sim N_f/N_c$
found in large $N_c$ expansion~\cite{Toublan}.
In Fig.~\ref{fig:critical_line} (right) we show our results fit. 
We checked the stability of the fit
by choosing different ranges in the chemical potential values.
Considering the entire range of values we find $t_2=0.89(4)$, discarding
the first value we find $t_2=0.89(5)$ and discarding the last one, 
$t_2=1.02(12)$. 

\section{The quark number density at imaginary chemical potential}

For a large homogeneous system, the pressure and its first derivatives, 
the quark number densities, are defined as
\begin{eqnarray}
  {p\over T^4}&=&{1\over{VT^3}}\ln{\cal Z}\\
  {n_{u,d,s}\over T^3}&=&{1\over{VT^2}}\frac{\partial\ln{\cal Z}}{\partial \mu_{u,d,s}}
\end{eqnarray}
where the partition function $\cal Z$ is a function of the volume $V$, 
temperature $T$,
quark masses $m_{u,d,s}$ and chemical potentials $\mu_{u,d,s}$.
Note that for imaginary chemical potential
the quark number density is purely imaginary~\cite{DElia2}.

\begin{figure}[htb]
  \centering
  \bigskip
  \includegraphics[width=5.5cm,height=4.5cm]{nud_ns.eps}$\qquad$
  \includegraphics[width=5.5cm,height=4.5cm]{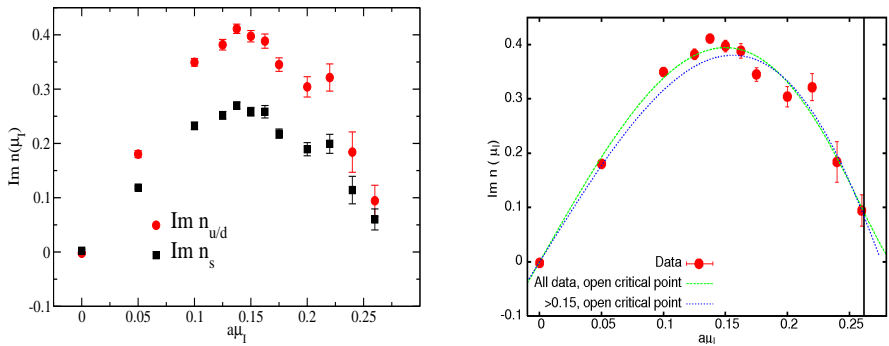}
  \caption{ (left) $n_l$ and $n_s$ in comparision; (right) quark number 
density $n_l$ at T=209.6 MeV fitted 
to the form predicted by a simple critical behavior at imaginary 
chemical potential. The vertical line indicates the Roberge Weiss 
transition.}
  \label{fig:critical_fit}
\end{figure}

We calculated the quark number density
at various values of the imaginary chemical potential $\mu = i \mu_I$.
In  Fig.~\ref{fig:critical_fit} (left) we show our results for the light,
$n_l$, $l=u,d$, and the strange quark number density, $n_s$, at the temperature
$T=209.6$ MeV.
We fitted our data to the ansatz 
\begin{equation}
  \Im \, n(\mu_I) = A \mu_I (c^2 - {\mu_I}^2)^e
\label{eq:critical_fit}
\end{equation}
which was suggested in \cite{DElia_fit}.
This ansatz takes into account that the quark number density
is an odd function of the chemical potential.
Moreover, if the exponent $e$ is less than 1, Eq.~\ref{eq:critical_fit}
leads to a singularity in the quark number susceptibility at $c$.
Eq.~\ref{eq:critical_fit} can also be derived from the singular part
of the $ln{\cal Z}$ in the vicinity of a critical point $c=\mu_I^c$, with
$e$ given by the critical exponent $\alpha$ as $e=1-\alpha$.
In Fig.~\ref{fig:critical_fit} (right) we show the results of the fit
to the light quark density obtained at a temperature of 209.6 MeV. 
A fit to our entire interval and with $c$
unconstrained gives $A=105(67)$, $c=0.285(14)$ and $e=1.30(28)$.
In the range $[0.15,0.26]$ of $a \mu_I$, we 
obtain $A=57(54)$, $c=0.276(15)$ and $e=1.06(36)$.
We thus find that $e$ is consistent with 1 
and substantially larger than the value $0.28(2)$ found
in~\cite{DElia_fit} while $c$ exceeds $\pi T/3$ slightly.
We interpret these results as to indicate that the regular
contributions to $ln{\cal Z}$ are dominating the number density at the temperatures
investigated (see also~\cite{Schmidt}).

One of the approaches to QCD at finite density is based on the Taylor
expansion of the pressure in terms of the quark chemical potentials
around a generic $\mu_0 = (\mu_{u,0}, \mu_{d,0}, \mu_{s,0})$,
\begin{equation}
  {p\over T^4}(\hat{\mu})=\sum_{k,l,n}c_{kln}(\hat{\mu}_u - \hat{\mu}_{u,0})^k
(\hat{\mu}_d - \hat{\mu}_{d,0})^l(\hat{\mu}_s - \hat{\mu}_{s,0})^n\nonumber
\end{equation}
with the abbreviation $\hat{\mu}_q = \mu_q/T$ and
the coefficients
\begin{equation}
  c_{kln} = \frac{1}{k!l!n!}\frac{\partial^k}{\partial\hat{\mu_u}^k}
\frac{\partial^l}{\partial\hat{\mu_d}^l}
\frac{\partial^n}{\partial\hat{\mu_s}^n}\left(p \over T^4\right)
\end{equation}
evaluated at $\mu_0$.
Correspondingly, the number density of e.g. the u quark is given by
\begin{equation}
  {n_u\over T^3}(\hat{\mu})=\sum_{k,l,n}kc_{kln}
(\hat{\mu}_u - \hat{\mu}_{u,0})^{k-1}
(\hat{\mu}_d - \hat{\mu}_{d,0})^l(\hat{\mu}_s - \hat{\mu}_{s,0})^n.
\end{equation}
Usually the coefficients are computed at $\mu_{q,0}=0$ but they could
also be calculated at imaginary values for the chemical potentials.
This is work in progress. 

\begin{figure}
  \centering
  \bigskip
  \includegraphics[width=.40\textwidth, angle=270]{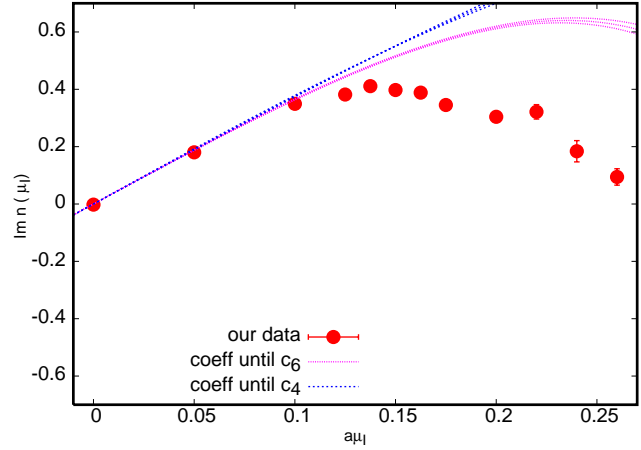}
  \caption{The light quark number density at 209.6 MeV calculated at imaginary
    chemical potentials (points) and from Taylor expansion (curves with
    error bands). The pink curve from \cite{Cheng:2008zh} takes into account 
    coefficients up to 6th order. 
    The blue curve has been obtained from our own computations of the
    coefficients until fourth order which have less statistics.
    }
  \label{fig:Imn}  
\end{figure}

In Fig.~\ref{fig:Imn} we compare our data for the quark number density 
calculated at imaginary chemical potentials with predictions from the 
Taylor expansion at a temperature of 209.6~MeV and at the same
lattice parameters. 
Up to $\mu_I/T \simeq 0.4$ there is practically no difference between 
the predictions obtained from Taylor expansion up to fourth and
6th order \cite{Cheng:2008zh}.
At larger values of $\mu_I$ the 6th order Taylor curve 
starts to deviate from the fourth order one and bends downwards, thus
qualitatively describing this trend of the data correctly.
Still, the errors arising from trucanting the Taylor series at sixth 
order become sizeable at $\mu_I/T \simeq 0.4$ at this temperature.

\section{Summary and conclusions}
We have studied QCD with 2+1 flavors at nonzero temperature and nonzero
chemical potential on a line of constant physics with the strange quark mass
adjusted to its physical value and the pion mass of about 220 MeV. 
The simulations have been carried out at imaginary chemical potentials. 
It turned out that the temperatures of up to $1.08 T_c$ at 
which the computations
were performed are below the endpoint $T_{RW}$ of a line of Roberge 
Weiss transitions. Instead, we could identify a crossover
line at which phase and modulus of the Polyakov loop as well as
the chiral condensate change their behavior. The curvature
of this line was estimated as $t_2=0.89(2)$
which is not inconsistent with the literature.
Furthemore.., we calculated quark number densities
at imaginary $\mu$ and compare our data with results
obtained via Taylor expansion method. Up to
$\mu_I/T \simeq 0.4$ good agreement was observed. 
 
\acknowledgments
This work has been supported in parts by the BMBF grant 06BI9001 and
the EU Integrated Infrastructure Initiative ``Hadron Physics 2''.
The numerical simulations have been performed on the apeNEXT at 
Bielefeld University.

\end{document}